\documentclass{article}
\usepackage[a4paper,margin=2cm]{geometry}
\usepackage[utf8]{inputenc}
\usepackage{amsmath}
\usepackage{color}
\usepackage{hyperref}
\usepackage{amsmath, amsfonts, amsthm, amssymb}
\usepackage{mathtools}
\usepackage{commath}
\usepackage{verbatim}
\usepackage[toc,page]{appendix}
\usepackage{stmaryrd}
\usepackage{fancyhdr}
\usepackage{bbm}
\usepackage{graphicx}
\usepackage{graphics}
\usepackage{tikz}
\usetikzlibrary{graphs}
\usetikzlibrary{arrows.meta}
\usepackage{caption}
\usepackage{subcaption}
\usepackage{booktabs}
\usepackage{array}
\usepackage{comment}
\usepackage{todonotes}
\usepackage{algorithm2e}
\usepackage{tcolorbox}
\usepackage{float}
\usepackage[noend]{algpseudocode}
\usepackage{tikz}
\usepackage{multirow,array}
\usetikzlibrary{matrix,chains,positioning,decorations.pathreplacing,arrows}
\usepackage[backend=bibtex, natbib=true, bibstyle=numeric, citestyle=numeric, doi=true, sorting=nyt, giveninits]{biblatex}
\usepackage{listings}
\usepackage{xcolor}
\usepackage{lipsum}
\definecolor{Darkgreen}{rgb}{0,0.4,0}
\newfam\msbfam

\newcommand{\Prob}{P}
\DeclareMathOperator{\ifr}{ifr}
\DeclareMathOperator{\AR}{AR}
\DeclareMathOperator{\Normal}{\mathcal{N}}

\theoremstyle{definition}

\usepackage{authblk}
\date{}

\title{Estimating the number of infections and the impact of non-pharmaceutical interventions on COVID-19 in European countries: technical description update}

\author{Seth Flaxman$^*$, Swapnil Mishra$^*$, Axel Gandy$^*$,  H Juliette T Unwin, Helen Coupland,  Thomas A Mellan, Harrison Zhu, Tresnia Berah, Jeffrey W Eaton, Pablo N P Guzman, Nora Schmit, Lucia Callizo, Imperial College COVID-19 Response Team, Charles Whittaker, Peter Winskill, Xiaoyue Xi, Azra Ghani, Christl A. Donnelly, Steven Riley, Lucy C Okell, Michaela A C Vollmer, Neil M. Ferguson and Samir Bhatt$^{*,1}$}

\date{Department of Infectious Disease Epidemiology, Imperial College London\\
Department of Mathematics, Imperial College London\\
WHO Collaborating Centre for Infectious Disease Modelling\\
MRC Centre for Global Infectious Disease Analytics\\
Abdul Latif Jameel Institute for Disease and Emergency Analytics, Imperial College London\\
Department of Statistics, University of Oxford\\
\ \\
$^*$Contributed equally. $^1$Correspondence: s.bhatt@imperial.ac.uk
}

\begin{document}
\maketitle
\newpage
\section*{Brief Introduction}
Following the emergence of a novel coronavirus (SARS-CoV-2) and its spread outside of China, Europe has experienced large epidemics. In response, many European countries have implemented unprecedented non-pharmaceutical interventions including case isolation, the closure of schools and universities, banning of mass gatherings and/or public events, and most recently, wide-scale social distancing including local and national lockdowns. 

In this technical update, we extend a semi-mechanistic Bayesian hierarchical model that infers the impact of these interventions and estimates the number of infections over time. Our methods assume that changes in the reproductive number – a measure of transmission - are an immediate response to these interventions being implemented rather than broader gradual changes in behaviour. Our model estimates these changes by calculating backwards from temporal data on observed to estimate the number of infections and rate of transmission that occurred several weeks prior, allowing for a probabilistic time lag between infection and death. 

In this update we extend our original model \citep{flaxman2020report13} to include (a) population saturation effects,  (b)  prior uncertainty on the infection fatality ratio, (c) a more balanced prior on intervention effects and (d) partial pooling of the lockdown intervention covariate. We also (e) included another 3 countries (Greece, the Netherlands and Portugal).

 The model code is available at \url{https://github.com/ImperialCollegeLondon/covid19model}.
We are now reporting the results of our updated model online at \url{https://mrc-ide.github.io/covid19estimates/}.
We estimated parameters jointly for all $M=14$ countries in a single hierarchical model. Inference is performed in the probabilistic programming language Stan \cite{carpenter2017stan} using an adaptive Hamiltonian Monte Carlo (HMC) sampler. 

\section{Model description}
We observe daily deaths $D_{t,m}$ for days $t\in\{1,\dots,n\}$ and countries $m\in\{1,\dots,M\}$. These daily deaths are modelled using a positive real-valued function $d_{t,m}=\mathbb{E}[D_{t,m}]$ that represents the expected number of deaths attributed to COVID-19. The daily deaths $D_{t,m}$ are assumed to follow a negative binomial distribution with mean $d_{t,m}$ and variance $d_{t,m}+\frac{d_{t,m}}{\phi}$, where ${\phi}$ follows a positive half normal distribution, i.e.  
\begin{align*}
    D_{t,m}&\sim \text{Negative Binomial}\left(d_{t,m}, d_{t,m}+\frac{d_{t,m}^2}{\psi}\right),\\
    \psi&\sim \Normal^+(0,5).
\end{align*}
Here,  $\Normal(\mu,\sigma)$ denotes a normal distribution with mean $\mu$ and standard deviation $\sigma$. We say that $X$ follows a positive half normal distribution $\Normal^+(\mu,\sigma)$ if $X\sim |Y|$, where $Y\sim\Normal(\mu,\sigma)$.

The expected number of deaths $d$ in a given country on a given day is a function of the number of infections $c$ occurring in previous days. Here infections are both symptomatic and asymptomatic.

At the beginning of the epidemic, deaths resulting from individuals infected abroad can bias estimates of the starting reproduction number $R_0$. To ensure that we are modelling deaths resulting from locally acquired infections we only include observed deaths from the day after a country has cumulatively observed $10$ deaths in our model. We tested the sensitivity of this parameter using Pareto smoothed importance sampling leave one out validation (PSIS-LOO) \cite{vehtari_practical_2017}  and found insignificant sensitivity around our choice.

To mechanistically link our function for deaths to our latent function for infected cases, we use a previously estimated COVID-19 infection-fatality-ratio $\ifr$ (probability of death given infection) together with a distribution of times from infection to death $\pi$. The $\ifr $ is derived from estimates presented in Verity et al \cite{verity_estimates_2020} which assumed homogeneous attack rates across age-groups. To better match estimates of attack rates by age generated using more detailed information on country and age-specific mixing patterns, we scale these estimates (the unadjusted $\ifr $, referred to here as $\ifr '$) in the following way as in previous work \cite{ferguson_impact_nodate}. Let $c_a$ be the number of infections generated in age-group $a$,  let $N_a$ be the underlying size of the population in that age group and let $\AR_a=c_a/N_a$ be the age-group-specific attack rate. The adjusted $\ifr $ is then given by:
$$\ifr _a=\frac{\AR_{50-59}}{\AR_a}\ifr _a',$$
where $\AR_{50-59}$ is the predicted attack-rate in the 50-59 year age-group after incorporating country-specific patterns of contact and mixing. This age-group was chosen as the reference as it had the lowest predicted level of underreporting in previous analyses of data from the Chinese epidemic\cite{verity_estimates_2020}. We obtained country-specific estimates of attack rate by age, $\AR_a$, for the $M$ countries in our analysis from a previous study which incorporates information on contact between individuals of different ages in countries across Europe\cite{walker_report_nodate}. We then obtained overall $\ifr $ estimates for each country adjusting for both demography and age-specific attack rates. Details of this calculation can be found in \cite{verity_estimates_2020}\cite{walker_report_nodate}. 

From the above, every country has a specific mean infection-fatality ratio $\ifr_m$. To incorporate the uncertainty inherent in this estimate we allow the $\ifr_m$ for every country to have some additional noise around the mean. Specifically we assume 
\begin{equation*}
\ifr^*_m\sim \ifr_m\cdot N(1,0.1).
\end{equation*}

Using estimated epidemiological information from previous studies\cite{verity_estimates_2020,walker_report_nodate}, we assume the distribution of times from infection to death  $\pi$ (infection-to-death) to be the sum of two independent random times: the incubation period (infection to onset of symptoms or infection-to-onset) distribution and the time between onset of symptoms and death (onset-to-death). The infection-to-onset distribution is Gamma distributed with mean $5.1$ days and coefficient of variation $0.86$. The onset-to-death distribution is also Gamma distributed with a mean of $17.8$ days and a coefficient of variation $0.45$.  The infection-to-death distribution is therefore given by:
\begin{align*}
\pi &\sim \text{Gamma}(5.1,0.86)+\text{Gamma}(17.8,0.45) .
\end{align*}

The expected number of deaths $d_{t,m}$, on a given day $t$, for country, $m$, is given by the following discrete sum:
\begin{equation*}
    d_{t,m} = \ifr^*_m\sum_{\tau=0}^{t-1} c_{\tau,m}\pi_{t-\tau},
\end{equation*}
where $c_{\tau,m}$ is the number of new infections on day $\tau$ in country $m$ and where $\pi$ is discretized via $\pi_{s}=\int_{s-0.5}^{s+0.5} \pi(\tau) d\tau$ for $s=2,3,...,$ and $\pi_{1} = \int_{0}^{1.5} \pi(\tau)d\tau$, where $\pi(\tau)$ is the density of $\pi$. 

The number of deaths today is the sum of the past infections weighted by their probability of death, where the probability of death depends on the number of days since infection and the country-specific infection-fatality-ratio (with some noise). 

The true number of infected individuals, $c$, is modelled using a discrete renewal process. This approach has been used in numerous previous studies and has a strong theoretical basis in stochastic individual-based counting processes such as Hawkes process, the Erland Susceptible-Exposed-Infected-Recovered model, and the Bellman-Harris process. To model the number of infections over time, we need to specify a generation distribution $g$ with density $g(\tau)$, (the time between when a person gets infected and when they subsequently infect another other people). The generation distribution is unknown, but we can approximate it by assuming it is the same as the serial interval distribution (time from onset to onset). We choose these to be Gamma distributed:
\begin{equation*}
    g\sim \text{Gamma}(6.5,0.62).
\end{equation*}
Given the generation distribution, the number of infections $c_{t,m}$ on a given day $t$, and country, $m$, is given by the following discrete convolution function:
\begin{align*}
    c_{t,m} &= S_{t,m}R_{t,m} \sum_{\tau=0}^{t-1}c_{\tau,m}g_{t-\tau}\\
     S_{t,m} &=1-\frac{\sum_{i=1}^{t-1}c_{i,m}}{N_m} 
\end{align*}
where, similar to the probability of death function, the generation distribution is discretized by $g_s=\int_{s-0.5}^{s+0.5} g(\tau) d\tau$ for $s=2,3,...,$ and $g_1 = \int_{0}^{1.5} g(\tau)d\tau$. The population of country $m$ is denoted by $N_m$ . We include the adjustment factor $S_{t,m} = 1-\frac{\sum_{i=1}^{t-1}c_{i,m}}{N_m}$  to account for the number of susceptible individuals left in the population: i.e even in the absence of interventions, herd immunity will reduce the number of daily infected. This of course assumes reinfection over the time horizon of our model is impossible.  We note here that we could include a factor in the serial interval accounting for individuals who die before they can infect others but given the infection-to-death distribution this factor is negligible and we have chosen to exclude it.

Infections today depend on the number of infections in the previous days, weighted by the discretized generation distribution. This weighting is then scaled by the country-specific time-varying reproduction number, $R_{t,m}$, that models the average number of secondary infections at a given time.
The functional form for the time-varying reproduction number was chosen to be as simple as possible to minimize the impact of strong prior assumptions: we use a piecewise constant function that scales $R_{t,m}$ from a baseline prior $R_{0,m}$ and is driven by known major non-pharmaceutical interventions occurring in different countries and times. 

We included 6 interventions, one of which is constructed from the other 5 interventions, which are timings of school and university closures ($k=1$), self-isolating if ill ($k=2$), banning of public events ($k=3$), any government intervention in place ($k=4$), implementing a partial or complete lockdown ($k=5$) and encouraging social distancing and isolation ($k=6$). We denote the indicator variable for intervention $k\in \{1,\dots,6\}$ by $I_{k,t,m}$, which is 1 if intervention $k$ is in place in country m at time t and 0 otherwise. The covariate “any government intervention” ($k=4$) indicates if any of the other 5 interventions are in effect, i.e. $I_{4,t,m}$ equals 1 at time t if any of the interventions $k \in \{1,2,3,5,6\}$ are in effect in country $m$ at time $t$ and equals 0 otherwise. Covariate 4 has the interpretation of indicating the onset of major government intervention. The effect of each intervention is assumed to be multiplicative. $R_{t,m}$ is therefore a function of the intervention indicators $I_{k,t,m}$ in place at time $t$ in country $m$:
\begin{equation*}
   R_{t,m} = R_{0,m}e^{-\sum_{k=1}^6 \alpha_k I_{k,t,m}-\beta_m I_{5,t,m}}
\end{equation*}
The exponential form was used to ensure positivity of the reproduction number, with $R_{0,m}$ constrained to be positive as it appears outside the exponential. 
The impacts $\alpha_k$ are shared between all $M$ countries and therefore they are informed by all available data. The intervention of a lockdown has another another country-specific, or partial pooled, random effect given by $\beta_m$. We justify the inclusion of this partially pooled effect as lockdown is the only identifiable parameter in a full pooled model.

The prior distribution for $R_{0,m}$ was chosen to be
\begin{align*}
R_{0,m}&\sim \Normal^+(3.28,|\kappa|)  \text{ with }
\kappa \sim \Normal^+(0,0.5),
\end{align*}where $\kappa$ is the same among all countries to share information about the variability of $R_{0,m}$. The value of $3.28$ was chosen based on a previous meta analysis looking at the basic reproductive number \cite{R0_paper}. 

The prior on the total reduction through one intervention (i.e. $\exp(- \alpha_k)$) which is not lockdown and on the reduction once all interventions in place, i.e.\ on $\exp(-\sum_{k=1}^6 \alpha_k)$, is displayed in the top row of Figure \ref{fig:priorreductionRt}.  Bottom row illustrates the prior on the effect of lockdown (i.e. $\exp\left(-\alpha_k-\beta_m\right))$ and full effect of all interventions together ($\exp\left(-\sum_{k=1}^{6}\alpha_k-\beta_m\right)$ bottom right). Details of the individual prior choices that result in this are below.

\begin{figure}
    \centering
\includegraphics[width=\linewidth]{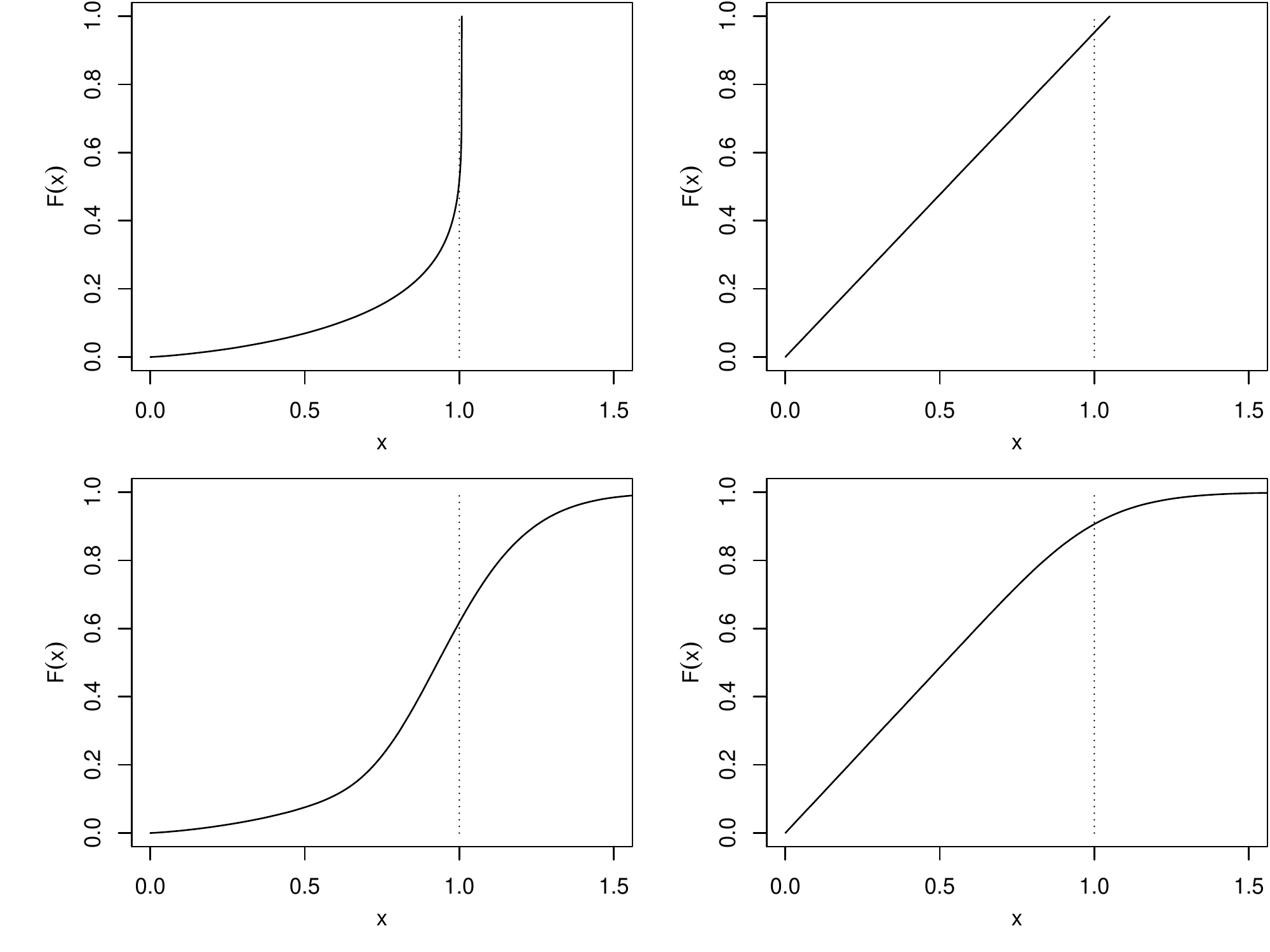}
    \caption{Cumulative distribution function of prior on total reduction through one intervention through the fixed effect ($\exp\left(-\alpha_k\right)$, top left), on the lockdown intervention ($\exp\left(-\alpha_k-\beta_m\right)$, bottom left) fixed effect through all interventions together ($\exp\left(-\sum_{k=1}^{6}\alpha_k\right)$, top right) and full effect of all interventions together ($\exp\left(-\sum_{k=1}^{6}\alpha_k-\beta_m\right)$ bottom right).}
    \label{fig:priorreductionRt}
\end{figure}


The impact of an intervention on $R_{t,m}$ is characterised by a set of parameters $\alpha_1,\dots,\alpha_6$, with independent prior distributions chosen to be 
\begin{equation*}
\alpha_k \sim \text{Gamma}(1/6,1) - \frac{\log(1.05)}{6},
\end{equation*}
i.e.\ the prior on each effect is a Gamma distribution with shape parameter 1/6 and scale parameter 1, shifted to allow for negative values.
This prior was chosen such that the probability that any individual intervention does not reduce $R_{t,m}$, i.e. $\Prob(\alpha_k<0)$, is about 48\% and such that the joint effect of $\alpha_1,\dots,\alpha_k$ on $R_{t,m}$ once all interventions are in-place (i.e. the distribution of $\exp(-\sum_{k=1}^6\alpha_k)$) is a uniform distribution on $[0,1.05]$. The intuition behind this prior is that it encodes our null belief that interventions could equally increase or decrease $R_t$, and the data should inform which.

The prior on the country-specific effects of lockdown $\beta_1,\ldots,\beta_M$ is given by
$$
\beta_1,\ldots,\beta_M\sim N\left(0,\gamma\right)\mathrm{\ where\ }\gamma\sim N^+\left(0,.2\right).
$$We only included such a country-specific random effect for lockdown, as the lockdown effect is the strongest in our analysis and as other interventions do not have identifiable effects.

We assume that seeding of new infections begins 30 days before the day after a country has cumulatively observed 10 deaths. From this date, we seed our model with 6 sequential days of an equal number of infections: $c_{1,m}=\dots =c_{6,m}\sim \text{Exponential}(\frac{1}{\tau})$, where $\tau \sim \text{Exponential}(0.03)$. These seed infections are inferred in our Bayesian posterior distribution.

We estimated parameters jointly for all countries in a single hierarchical model. Fitting was done in the probabilistic programming language Stan\cite{carpenter2017stan} using an adaptive Hamiltonian Monte Carlo (HMC) sampler. 

\printbibliography

\end{document}